\documentclass[nohyper,12pt,letterpaper]{JHEP3}
\usepackage{epsfig}

\newcommand{\myfig}[3]{\begin{figure}[ht]
\begin{center}
\leavevmode \epsfxsize=#2cm \epsfbox{#1}
\end{center}
\caption{#3} \label{fig:#1}
\end{figure}}

\author{Robert de Mello Koch$^{1,2}$ Tanay K. Dey$^{1}$, Norman Ives$^{1}$ and Michael Stephanou$^{1}$\\
\qquad \\
$^{1}$ National Institute for Theoretical Physics,\\
Department of Physics and Centre for Theoretical Physics,\\ 
University of the Witwatersrand,\\ 
Wits, 2050,\\ 
South Africa\\
\qquad\\
$^{2}$Stellenbosch Institute for Advanced Studies,\\
Stellenbosch,\\
South Africa\\
\qquad\\
E-mail: \email{robert@neo.phys.wits.ac.za, Tanay.Dey@wits.ac.za, Norman.Ives@students.wits.ac.za, Michael.Stephanou@students.wits.ac.za}}

\abstract{The problem of computing the anomalous dimensions of a class of (nearly) half-BPS operators with a 
large ${\cal R}$-charge is reduced to the problem of diagonalizing a Cuntz oscillator chain. Due to the large
dimension of the operators we consider, non-planar corrections must be summed to correctly construct the Cuntz
oscillator dynamics. These non-planar corrections do not represent quantum corrections in the dual gravitational
theory, but rather, they account for the backreaction from the heavy operator whose dimension we study. Non-planar
corrections accounting for quantum corrections seem to spoil integrability, in general. It is interesting to ask 
if non-planar corrections that account for the backreaction also spoil integrability. We find a limit in which
our Cuntz chain continues to admit extra conserved charges suggesting that integrability might survive. 
}

\preprint{WITS-CTP-044}

\title{Hints of Integrability Beyond the Planar Limit: Nontrivial Backgrounds}


\def \Tr{\mbox{Tr\,}}

\begin{document}

\section{Introduction}

There is by now an impressive body of work suggesting that {\sl planar} ${\cal N} = 4$ super Yang-Mills theory is 
{\sl exactly integrable}. This would be very fortunate indeed, since it would mean the problem of computing the 
spectrum of all possible scaling dimensions of the gauge theory can be solved exactly,
in the large $N$ limit, by employing a Bethe ansatz. 
This has been established for the complete set of possible operators at one loop in\cite{Minahan:2002ve,Beisert:2003yb}
and to two and three loops in the su$(2)$ sector\cite{Beisert:2003tq}. Given these results it is natural to guess
that integrability extends to all orders in perturbation theory and perhaps even to the non-perturbative level\cite{Beisert:2003tq}. 
There is now mounting evidence that this guess is correct
\cite{Beisert:2003ys,Serban:2004jf,Beisert:2004hm,Eden:2004ua,Staudacher:2004tk,Kazakov:2004nh,Beisert:2005tm,Zwiebel:2005er,Beisert:2005fw,Beisert:2005wm,Eden:2006rx,Beisert:2006ib,Beisert:2006ez}.

The idea of spin chain parity played a central role in the discovery of the planar two and three loop integrability\cite{Beisert:2003tq}.
Acting on a single trace operator, parity simply reverses the order of fields inside the trace. The dilatation operator commutes
with parity, so that as we would expect, the dilatation eigenstates are also parity eigenstates. In addition, \cite{Beisert:2003tq} found
that eigenstates with opposite parity were degenerate - this was quite unexpected. This degeneracy can be explained by the existence of
a higher conserved charge ($U_2$ in the notation of \cite{Beisert:2003tq}) that commutes with the dilatation operator 
but anticommutes with parity. Its also possible (and useful) to put this 
argument on its head: one can interpret the degeneracy of states with opposite parity as evidence for the existence of further conserved
quantities. When non-planar corrections are taken into account, parity remains a good quantum number but the degeneracy is lifted
(see \cite{Beisert:2003tq} for a discussion of the ${\cal N}=4$ super Yang-Mills case and \cite{Kristjansen:2008ib} for a very
relevant discussion in the context of the ABJM theory). Although
this only proves that the standard construction of conserved charges does not work away from the strict planar limit, it does suggests
that integrability might not survive away from this limit. Clearly an important question is

{\vskip 0.1cm}

{\sl Does integrability survive non-planar corrections?}

{\vskip 0.1cm}

\noindent
In this article we will explicitly describe a situation in which we do sum (in fact, an infinite number of) 
non-planar corrections. Further, we collect some evidence that the resulting dynamics remains integrable. This
suggests that, at least in certain situations, the answer to the above question is positive. 

The non-planar corrections that we consider arise because we are interested in computing the anomalous dimensions of operators whose classical
dimension is of order $N^2$. The usual $1/N$ suppression of non-planar diagrams is, in this case, overpowered by huge combinatoric
factors\cite{Balasubramanian:2001nh}. In a series of articles\cite{Koch:2008ah,Koch:2008cm,Koch:2009jc}, building on the earlier
works\cite{Corley:2001zk,Corley:2002mj,de Mello Koch:2004ws}, we have developed techniques to systematically study these operators.
The specific operators we study are spelt out in detail in section 2.1. In section 2.2 we give the dilatation operator to two loops.
To obtain this result, ribbon diagrams with arbitrarily large genus are summed. In section 2.3 we study the action of our dilatation operator.
The action of this dilatation operator is {\sl not easily formulated in a spin chain language because, nonplanar corrections allow the number 
of fields within each trace to change}. This translates into a spin chain with a variable lattice length. A convenient reformulation
of the problem, in terms of a Cuntz lattice, was given in \cite{Berenstein:2005fa}. In the reformulation, particles (described by Cuntz
oscillators) hop on a lattice of fixed size. The fact that the size of the spin chain lattice was dynamical now translates into the fact
that the total number of particles populating the Cuntz lattice is dynamical. We give the Cuntz oscillator description of the dilatation
operator for the class of operators we study in section 2.4. In section 3 we start to look for signs that our Cuntz chain is integrable.
To start, we rewrite the spin chain (in the su$(2)$ sector) of \cite{Minahan:2002ve,Beisert:2003yb,Beisert:2003tq} in the Cuntz oscillator 
language. In particular, we write the conserved charge $U_2$ of\cite{Beisert:2003tq} in terms of Cuntz oscillators. We have verified, using 
this expression for $U_2$, that $U_2$ does not commute with the Cuntz chain Hamiltonian corresponding to the annulus geometry. Another way
explore the integrability of the original model is to study the semi-classical limit of the spin chain, which can be matched to the low energy limit of the
principal chiral model. It is known that the principal chiral model is integrable. We can show that the Cuntz chain corresponding to the spin 
chain of \cite{Minahan:2002ve,Beisert:2003yb,Beisert:2003tq} is indeed equivalent to the low energy limit of the principal chiral model - the
spin chain and the Cuntz chain simply correspond to different choices of gauge. We give the explicit form of the gauge transformation relating the two in section 3.2.
In section 3.3 we study the large $M$ limit of our Hamiltonian and argue that we can indeed write down higher conserved charges. This suggests
that integrability survives in this limit. In section 4 we discuss our results.

\section{Two Loop Cuntz Chain of the LLM Background}

${\cal N}=4$ super Yang-Mills theory has 6 scalars $\phi_i$ transforming in the adjoint of the gauge group and in the 
${\bf 6}$ of the $SU(4)_{\cal R}$ symmetry. We shall use the complex combinations
\begin{equation} 
Z=\phi_1+i\phi_2,\qquad Y=\phi_3+i\phi_4,\qquad X=\phi_5+i\phi_6,
\end{equation}
in what follows. All operators that we study are built using only $Z$ and $Y$; they belong to the su$(2)$ sector of the theory. 
Many of the expressions that we write involve traces over $Z$s, $Y$s and derivatives of them. To avoid confusion, we will now
spell out the index structure of a few expressions
\begin{equation} 
\Tr\left( Z{\partial\over\partial Z}\right) = Z_{ij}{\partial\over\partial Z_{ij}}\, ,
\end{equation}
\begin{equation} 
\Tr \left( ZY{\partial\over\partial Z}{\partial\over\partial Y}\right) =
Z_{ij}Y_{jk}{\partial\over\partial Z_{lk}}{\partial\over\partial Y_{il}}\, .
\end{equation}

\subsection{Annulus Background}

Schur polynomials provide a very convenient reorganization of the half-BPS sector of the ${\cal N}=4$ super 
Yang-Mills theory. This is due to the fact that their two point function is known to all orders in 
${1\over N}$\cite{Corley:2001zk,Corley:2002mj} and that they satisfy a product rule allowing computation of exact $n$-point 
correlators using only two-point functions\footnote{There are generalizations of these results to multimatrix models\cite{mm}.
These results will be very relevant for studies of backgrounds that preserve less supersymmetry.}. The half-BPS sector of the theory can be reduced to the dynamics
of the eigenvalues of $Z$, which is the dynamics of $N$ non interacting fermions
in an external harmonic oscillator potential\cite{Corley:2001zk,Berenstein:2004kk}.
The half-BPS sector of operators with ${\cal R}$ charge of order 
$N^2$, are dual to solutions of type IIB supergravity - the LLM geometries\cite{Lin:2004nb}. For a careful
discussion of precisely what aspects of the eigenvalue dynamics the supergravity captures and vice versa, 
see \cite{Skenderis:2007yb}. For a recent discussion of the macroscopic description of the dual geometry see \cite{Simon:2009mf}.
It is by now well known that the space of the LLM geometries is given by a black 
and white coloring of a two dimensional plane\cite{Lin:2004nb,Balasubramanian:2005mg}; this colored plane is isomorphic to the fermionic phase
space of the eigenvalues of $Z$. The Schur polynomials correspond to geometries 
that are invariant under rotations in this plane.

The Schur polynomials are labeled by Young diagrams. In what follows, we will be interested in $\chi_B(Z)$ with $B$
a Young diagram that has $N$ rows and $M$ columns. We take $M$ to be of order $N$. Note that we can also express
\begin{equation} 
\chi_B(Z) =\left(\det (Z)\right)^M\, .
\end{equation}
This implies that
\begin{equation} 
{\partial\over\partial Z_{ij}}\chi_B(Z)= M(Z^{-1})_{ji}\chi_B(Z)\, ,
\end{equation}
a formula that we will make good use of below. The coloring describing the dual LLM geometry is a black annulus. The inner
white disk has an area of ${M\pi\over N}$ whilst the black annulus itself has an area of $\pi$ in units that assign an area
of ${\pi\over N}$ to each fermion state in phase space.

\subsection{Two Loop Effective Dilatation Operator}

The two loop dilatation operator, in the su$(2)$ sector, has been computed in \cite{Beisert:2003tq}. Using the conventions of
\cite{Beisert:2003tq}, the dilation operator can be expanded as
\begin{equation}
D=\sum_{k=0}^\infty \left( {g_{YM}^2\over 16\pi^2}\right)^k D_{2k}=
\sum_{k=0}^\infty g^{2k} D_{2k}\, ,
\end{equation}
where the tree level, one loop and two loop contributions are
\begin{equation}
D_0 = \Tr\left(Z {\partial\over \partial Z}\right)+\Tr\left(Y {\partial\over \partial Y}\right)\, ,
\end{equation}
\begin{equation}
D_2 = -2 : \Tr \left( \left[ Z,Y\right]\left[{\partial\over \partial Z},{\partial\over \partial Y}\right]\right) :\, ,
\end{equation}
$$
D_4 = -2 :\Tr \left(\left[\left[Y,Z\right],{\partial\over \partial Z}\right]
\left[\left[{\partial\over \partial Y},{\partial\over \partial Z}\right],Z\right]\right):
$$
$$
-2 :\Tr \left(\left[\left[Y,Z\right],{\partial\over\partial Y}\right]
\left[\left[{\partial\over\partial Y},{\partial\over\partial Z}\right],Y\right]\right):
$$
\begin{equation}
-2 :\Tr \left(\left[\left[Y,Z\right],T^a \right]
\left[\left[{\partial\over\partial Y},{\partial\over\partial Z}\right],T^a\right]\right):\, .
\end{equation}
The normal ordering symbols here indicate that derivatives within the normal ordering symbols do not act on fields inside the normal
ordering symbols.

We allow this dilatation operator to act on an operator built mainly from $Z$s with a few ``impurities'' $=Y$s added.
For most of our study we explicitly display formulas for operators with two or three impurities. Our final formulas are
however, completely general, covering the case that we have $O(1)$ impurities. Adapting the notation of \cite{Beisert:2003tq} we define
\begin{equation} 
{\cal O}_B (p,J_0,J_1,...,J_k)\equiv \chi_B (Z){\cal O}_p^{J_0;J_1,...,J_k} \equiv \chi_B (Z)\Tr (YZ^pYZ^{J_0-p})\prod_{i=1}^k \Tr Z^{J_i}\, ,
\end{equation}
\begin{equation} 
{\cal Q}_B (J_0,J_1,J_2,...,J_k)\equiv \chi_B (Z){\cal Q}^{J_0,J_1;J_2,...,J_k} \equiv \chi_B (Z)\Tr (YZ^{J_0})\Tr(YZ^{J_1})\prod_{i=2}^k \Tr Z^{J_i}\, ,
\end{equation}
where $\chi_B (Z)$ is the operator creating the background, which was defined in the previous section. Our
strategy is to define an effective dilatation operator $D_{\rm eff}$ as
\begin{equation} 
D \left(\chi_B (Z){\cal O}_p^{J_0;J_1,...,J_k} \right)=\chi_B(Z) D_{\rm eff}{\cal O}_p^{J_0;J_1,...,J_k}\, .
\end{equation}
Diagonalizing the action of $D_{\rm eff}$ on the gauge invariant operators ${\cal O}_p^{J_0;J_1,...,J_k}$ is clearly equivalent to diagonalizing
the action of $D$ on ${\cal O}_B (p,J_0,J_1,...,J_k)$. It is natural to interpret $D_{\rm eff}$ as the dilatation operator for
the LLM background, that is, for the theory that is deformed by the insertion of $\chi_B(Z)\chi_B(Z^\dagger )$
in the path integral. Notice that we can write
\begin{equation} 
D_{\rm eff}={1\over \chi_B(Z)} D \chi_B(Z)\, .
\end{equation}
This formula remains correct even after replacing $\chi_B(Z)$ by any other operator creating the background, which depends only on $Z$.
Ultimately, we will restrict ourselves to the large $M+N$ limit. To capture this limit, as explained in \cite{Koch:2009jc} one needs to
resum an infinite number of nonplanar diagrams; ribbon diagrams with arbitrarily large genus are summed.
{\sl This limit is certainly not the planar limit of ${\cal N}=4$ super Yang Mills theory.}

The crucial observation needed in the computation of $D_{\rm eff}$ is that
\begin{equation} 
{\partial\over\partial Z_{ij}}\left(\chi_B (Z){\cal O}_p^{J_0;J_1,...,J_k} \right)
=\chi_B (Z)\left({\partial\over\partial Z_{ij}}+M(Z^{-1})_{ji}\right){\cal O}_p^{J_0;J_1,...,J_k}\, .
\end{equation}
Repeated application of this formula gives
\begin{equation} 
D_{\rm 0\,\, eff}= D_0 + MN\, ,
\end{equation}
\begin{equation}
D_{\rm 2\,\, eff}= D_2 -2 M \Tr \left( \left(ZYZ^{-1}+Z^{-1}YZ-2Y\right){\partial\over \partial Y}\right) \, ,
\label{d2eff}
\end{equation}
\begin{eqnarray}
\nonumber
D_{\rm 4\,\, eff}&=& D_4 + 4NM \Tr \left( \left(ZYZ^{-1}+Z^{-1}YZ-2Y\right){\partial\over \partial Y}\right) \label{d4eff} \\
\nonumber
&+& 2M:\Tr \left[Z,Y\right]\left[Z^{-1},\left[Z,\left[{\partial\over \partial Z},{\partial\over \partial Y}\right]\right]\right]:
+2M:\Tr \left[Z,Y\right]\left[{\partial\over \partial Z},\left[Z,\left[Z^{-1},{\partial\over\partial Y}\right]\right]\right]: \\
\nonumber
&+&2M:\Tr \left[Z,Y\right]\left[{\partial\over\partial Y},\left[Y,\left[Z^{-1},{\partial\over\partial Y}\right]\right]\right]:\\
&-&2M^2 \Tr\left(\left(Z^2 Y Z^{-2}-4ZYZ^{-1}+6Y-4Z^{-1}YZ+Z^{-2}YZ^2\right){\partial\over\partial Y}\right)
\end{eqnarray}
for the tree level, one loop and two loop contributions to $D_{\rm eff}$. The formula for $ D_{\rm 0\,\, eff}$ has a straight forward
interpretation - the dimension of the gauge invariant operator ${\cal O}_p^{J_0;J_1,...,J_k}$ is shifted by $MN$ due to the presence of the background, which 
has dimension $MN$. The answer for $D_{\rm 2\,\, eff}$ has already been obtained and discussed in \cite{Chen:2007gh,Koch:2008ah,Koch:2008cm}.

\subsection{Action of the Two Loop Effective Dilatation Operator}

A useful observation made in \cite{Beisert:2003tq}, is that, when acting with $g^2 D_2 + g^4 D_4$ on the generic two impurity gauge invariant operators 
${\cal O}_B(p,J_0,J_1,...,J_k)$ and ${\cal Q}_B(J_0,J_1;J_2,...,J_k)$ operators of type ${\cal Q}_B(J_0,J_1;J_2,...,J_k)$ are never
produced. This is easy to understand: acting with $g^2 D_2 + g^4 D_4$ always inserts a commutator $\big[Y,Z\big]$ into a trace; this trace 
vanishes unless it contains another $Y$. This observation generalizes: when acting with $g^2 D_{\rm 2\,\, eff}+ g^4 D_{\rm 4\,\, eff}$ 
on the generic two impurity gauge invariant operators ${\cal O}_p^{J_0;J_1,...,J_k}$ and ${\cal Q}^{J_0,J_1;J_2,...,J_k}$ operators of type 
${\cal Q}^{J_0,J_1;J_2,...,J_k}$ are never produced. This follows because both
\begin{equation}
\Tr \left( \left(ZYZ^{-1}+Z^{-1}YZ-2Y\right){\partial\over \partial Y}\right) \
\end{equation}
and
\begin{equation}
\Tr\left(\left(Z^2 Y Z^{-2}-4ZYZ^{-1}+6Y-4Z^{-1}YZ+Z^{-2}YZ^2\right){\partial\over\partial Y}\right) 
\end{equation}
annihilate gauge invariant operators containing a single $Y$ and because the remaining terms in $g^2 D_{\rm 2\,\, eff}+ g^4 D_{\rm 4\,\, eff}$
always insert a $\big[Y,Z\big]$ into a trace. In what follows we will study the anomalous dimensions of the operators ${\cal O}_B (p,J_0,J_1,...,J_k)$.
Clearly, from the observation we just made, we do not need to consider the operators ${\cal Q}_B (J_0,J_1,J_2,...,J_k)$ to do this.

Consider the action of $D_{\rm 2\,\, eff}$ on ${\cal O}_p^{J_0;J_1,...,J_k}$. Following \cite{Beisert:2003tq} we can break $D_{\rm 2\,\, eff}$
into three pieces
\begin{equation} 
D_{\rm 2\,\, eff}= D_{\rm 2,0\,\, eff}+D_{\rm 2,+\,\, eff}+D_{\rm 2,-\,\, eff}\, ,
\end{equation}
where $D_{\rm 2,0\,\, eff}$ preserves the number of traces in ${\cal O}_p^{J_0;J_1,...,J_k}$, $D_{\rm 2,+\,\, eff}$ increases (by 1) the number
of traces and $D_{\rm 2,-\,\, eff}$ decreases (by 1) the number of traces. From (\ref{d2eff}) it is clear that the additional term proportional
to $M$ can only contribute to $D_{\rm 2,0\,\, eff}$. The terms $D_{\rm 2,+\,\, eff}$ and $D_{\rm 2,-\,\, eff}$ which involve gauge invariant operator splitting and
joining will not contribute at the leading order; they will be important when computing, for example, 
subleading corrections to the leading $M+N$ limit. The additional
contributions proportional to $M$ in (\ref{d2eff}) have an important effect: they imply that $p$ and $J_0-p$ can be negative. Thus,
we need to consider gauge invariant operators in which we populate the two ``gaps between the $Y$s'' with both positive and negative powers of $Z$. This implies that
$p$ in ${\cal O}_p^{J_0;J_1,...,J_k}$ is completely unrestricted.

It is easy to write down an exact expression for the action of $D_{\rm 2\,\, eff}$
$$ 
D_{\rm 2,0\,\, eff}{\cal O}_p^{J_0;J_1,...,J_k} = -4 A_1 \left({\cal O}_{p+1}^{J_0;J_1,...,J_k}\right.
$$
$$
\left. -{\cal O}_p^{J_0;J_1,...,J_k}\right)-4 A_2 \left({\cal O}_{p-1}^{J_0;J_1,...,J_k}\right.
$$
\begin{equation}
\left. -{\cal O}_p^{J_0;J_1,...,J_k}\right),
\label{d2eqn}
\end{equation}
$$ D_{\rm 2,+\,\, eff}{\cal O}_p^{J_0;J_1,...,J_k} = 4\sum_{J_{k+1}=1}^{p-1}
\Big({\cal O}_{p-J_{k+1}}^{J_0-J_{k+1};J_1,...,J_k,J_{k+1}}$$
$$-{\cal O}_{p-J_{k+1}-1}^{J_0-J_{k+1};J_1,...,J_k,J_{k+1}}\Big)
-4\sum_{J_{k+1}=1}^{J_0-p-1}
\Big({\cal O}_{p+1}^{J_0-J_{k+1};J_1,...,J_k,J_{k+1}}$$
\begin{equation}
-{\cal O}_{p}^{J_0-J_{k+1};J_1,...,J_k,J_{k+1}}\Big)\, ,
\end{equation}
$$ D_{\rm 2,-\,\, eff}{\cal O}_p^{J_0;J_1,...,J_k} = 4\sum_{i=1}^{k}
J_i \Big({\cal O}_{p+J_i}^{J_0+J_{i};J_1,..,\hat{J}_i,..,J_k}$$
$$-{\cal O}_{p+J_i-1}^{J_0+J_i;J_1,..,\hat{J}_i,..,J_k}\Big)
-4\sum_{i=1}^{k}J_i
\Big({\cal O}_{p+1}^{J_0+J_i;J_1,..,\hat{J}_i,..,J_k}$$
\begin{equation}
-{\cal O}_{p}^{J_0+J_{i};J_1,..,\hat{J}_i,..,J_k}\Big)\, ,
\end{equation}
where in the last expression hatted variables are removed from the argument of ${\cal O}$ and
\begin{eqnarray}
A_1 &=& M+N\qquad J_0>p\nonumber\\
&=& M\qquad {\rm otherwise} 
\end{eqnarray}
\begin{eqnarray}
A_2 &=& M+N\qquad p>0\nonumber\\
&=& M\qquad {\rm otherwise} \, .
\end{eqnarray}

Again motivated by \cite{Beisert:2003tq} we can write
\begin{equation} 
D_{\rm 4\,\, eff}=-{1\over 4}(D_{\rm 2\,\, eff})^2 + \delta D_{\rm 4\,\, eff} 
\end{equation}
where
$$ \delta D_{\rm 4\,\, eff} = 2:\Tr \left[ Z,Y\right]\left[{d\over dY},\left[Y,\left[{d\over dZ},{d\over dY}\right]\right]\right]:
+ 2M:\Tr \left[ Z,Y\right]\left[{d\over dY},\left[Y,\left[Z^{-1},{d\over dY}\right]\right]\right]:\, .
$$
We can decompose $\delta D_{\rm 4\,\, eff}$ as
\begin{equation} 
\delta D_{\rm 4\,\, eff} = \delta D_{\rm 4,0\,\, eff}
+\delta D_{\rm 4,+\,\, eff}+\delta D_{\rm 4,-\,\, eff}+\delta D_{\rm 4,+-\,\, eff}
+\delta D_{\rm 4,--\,\, eff}+\delta D_{\rm 4,++\,\, eff}\, .
\end{equation}
The number of pluses/minuses in the subscripts on the right hand side of this last expression tell us how many traces
are added/removed by the action of that particular term. $\delta D_{\rm 4,+-\,\, eff}$ adds and removes a trace and
hence it comes from summing higher genus ribbon diagrams which contribute to the trace conserving 
piece of $\delta D_{\rm 4\,\, eff}$. It is again easy to write 
down exact expressions for the action of these terms. These expressions are given in Appendix C.

\subsection{Leading $M+N$ Limit}

To extract the leading terms at large $N$ (recall that we take $M$ to be of order $N$) we need to rewrite the action of $D_{\rm 2\,\, eff}$ obtained
in the last subsection in terms of normalized gauge invariant operators, that is, gauge invariant operators that have a suitably normalized two point function. The relevant two point correlators
have been computed in Appendix A. The result is most easily written in terms of a Cuntz oscillator chain. We will now focus on gauge invariant operators that have $k=0$.
If we relax the restriction to two impurities (which we do from now on), the gauge invariant operators we study have the form
\begin{equation} 
{\cal O}_B (\{ p\} ,J_0) \equiv \chi_B (Z)\Tr (YZ^{p_1}YZ^{p_2}\cdots YZ^{p_n})\, ,\qquad \sum_{i=1}^n p_i = J_0\, .
\end{equation}
To translate this gauge invariant operator into a Cuntz chain state, we associate a site of the Cuntz chain with each of the gaps between the $Y$s. The above gauge invariant operator
defines a state in a Cuntz chain with $n$ sites. Further, the number of $Z$s in each site gives the occupation number of that site. Finally, we
require that operators with a normalized (free field) two point function map into normalized Cuntz chain states. This last point deserves a few 
comments. The spacetime dependence of the free field correlators we compute is trivially determined by the bare dimension of the operator. Thus
we can compute all correlators in zero dimensions. In this zero dimensional model, each two point function is just a number. An operator with a
normalized two point function is one for which this number is one. This map between two point functions and the norm of states is of course nothing
but the usual state operator correspondence.

The dilatation operator allows the $Z$s to hop between sites of the Cuntz chain. At leading order in large $M+N$, the action of $D_{\rm 2\, eff}$
is given entirely by $D_{\rm 2,0\, eff}$. Using the correspondence given in the last equation of Appendix A, the operator equation (\ref{d2eqn})
can be translated into an equation for the action of $D_{\rm 2\, eff}$ on the Cuntz lattice. If all Cuntz lattice occupation numbers are
positive then, since $D_{\rm 2\, eff}$ lowers each occupation number by at most 1, acting with $D_{\rm 2\, eff}$ can't change the value $p_- = 0$
where $p_-$ is negative the sum of all the negative occupation numbers. This implies that all gauge invariant operators have the same leading two point function
and hence, when acting on a Cuntz lattice with, for example, two sites ($\lambda = g^2 N$)
\begin{equation} 
g^2 D_{\rm 2,0\,\, eff}|\{ p_1, p_2 \}\rangle = -4\lambda {N+M\over N} \left(
|\{ p_1+1 ,p_2-1 \}\rangle -2|\{ p_1,p_2 \}\rangle+|\{ p_1-1,p_2+1 \}\rangle\right)\, .
\end{equation}
Similarly, if all Cuntz lattice occupation numbers are negative then, since $D_{\rm 2\, eff}$ raises each occupation number by at most 1, acting 
with $D_{\rm 2\, eff}$ again can't change the value $p_- = \sum_i p_i$. Again all gauge invariant operators have the same leading two point function
and hence, when acting on a Cuntz lattice with, for example, two sites
\begin{equation}
g^2 D_{\rm 2,0\,\, eff}|\{ p_1, p_2 \}\rangle = -4\lambda {M\over N} \left(
|\{ p_1+1 ,p_2-1 \}\rangle -2|\{ p_1,p_2 \}\rangle+|\{ p_1-1,p_2+1 \}\rangle\right)\, .
\end{equation}
Finally, consider for example the case that $D_{\rm 2,0\,\, eff}$ acts on a lattice with two sites and occupation numbers $p_1=0$, $p_2=2$.
In this case, the normalization of the gauge invariant operators is not the same: three terms have $p_-=0$ and one has $p_-=1$. Taking this into account gives
$$ g^2 D_{\rm 2,0\,\, eff}|\{ 0, 2 \}\rangle = -4\lambda {M+N\over N} \left(
|\{ 1 ,1 \}\rangle -|\{ 0,2 \}\rangle\right)$$
\begin{equation}
-4\lambda {M\over N}\left( {\sqrt{M+N}\over\sqrt{M}}|\{ -1,3 \}\rangle- |\{ 0,2 \}\rangle\right)\, .
\end{equation}
There is a nice convenient way to summarize these results\cite{Chen:2007gh,Koch:2008ah,Koch:2008cm}. We will introduce Cuntz oscillators which satisfy the algebra
(we associate one of these oscillators to each site of the chain)
\begin{equation}
a^\dagger a ={M\over N} +\theta (\hat{n}+1) -|0\rangle \langle 0|\, ,\qquad a a^\dagger ={M\over N} +\theta (\hat{n}+1)\, ,
\end{equation}
with $\hat{n}$ the number operator.
Notice that when $M=0$ we have only positive occupation numbers so that the above relations reduce to the usual ones
\begin{equation}
a^\dagger a = 1 -|0\rangle \langle 0|\, ,\qquad a a^\dagger = 1 \, .
\end{equation}
We can also define these oscillators by giving their action on states of good particle number
\begin{equation}
a|n\rangle =\sqrt{1+{M\over N}} |n-1\rangle \qquad n>0 
\end{equation}
\begin{equation} 
a|n\rangle =\sqrt{M\over N} |n-1\rangle \qquad n\le 0\, . \
\end{equation}
In terms of these Cuntz oscillators we have
\begin{equation} 
g^2 D_{\rm 2\,\, eff} = 2 \lambda\sum_{l=1}^L (a_l^\dagger - a_{l+1}^\dagger)(a_l - a_{l+1})\, .
\end{equation}

There is a rather direct way to extract (part of the) geometry of the dual LLM solution from this Cuntz oscillator 
description\cite{Chen:2007gh,Koch:2008ah,Koch:2008cm}. To see this, consider the coherent state
\begin{equation} 
|z\rangle = \sum_{n=-\infty}^0 \left({N\over M}\right)^{n\over 2}z^n |n\rangle +\sum_{n=1}^\infty 
\left( {N\over M+N} \right)^{n\over 2} z^n |n\rangle\, .
\end{equation}
The norm of this state
\begin{equation}
\langle z|z\rangle = \sum^{\infty}_{n=0} {M^n \over N^n |z|^{2n}}+\sum_{n=1}^\infty {N^n |z|^{2n}\over (M+N)^{n}}
\end{equation}
is only finite if ${M\over N}\le |z|^2\le {M+N\over N}$, that is, $|z|^2$ must lie within the annulus. Clearly $z$ is a complex coordinate for the 
LLM plane.

This one loop result is intriguing:
the effect of the background has been completely accounted for by simply modifying the Cuntz oscillator algebra. This is a
remarkably simple change. It is natural to ask

{\vskip 0.1cm}

{\sl Is the net effect of the background (even at higher loops) completely accounted for, by simply modifying the Cuntz oscillator algebra?}

{\vskip 0.1cm}

\noindent
We do not have a complete answer to this question. Computing the form of the two loop answer ($D_{\rm 4\,\, eff}$) is already a rather
involved task. We have however studied this question for two sites. In this case
\begin{equation}
g^4 D_{\rm 4\,\, eff} = -{1\over 4} (g^2 D_{\rm 2\,\, eff})^2 +g^4\delta D_{\rm 4\,\, eff}
\end{equation}
where
$$
g^4\delta D_{\rm 4\,\, eff} =4\lambda^2\sum_{l=1}^2\left( 
 a_l^\dagger\big[ a_{l+1},a_{l+1}^\dagger\big]a_{l+1}
+a_l^\dagger a_{l+1}\big[ a_{l},a_{l}^\dagger\big]\right.
$$
\begin{equation}
\left.
-a_l^\dagger a_l \big[ a_{l+1},a_{l+1}^\dagger\big]
-a_l^\dagger\big[ a_{l},a_{l}^\dagger\big]a_{l}\right)\, .
\end{equation}
Thus, at two loops for a Cuntz chain with two sites we find that, again, the net effect of the background is 
to modify the Cuntz oscillator algebra.

\section{Conservation Laws}

In the last section we have explained how to extract an effective dilatation operator. Diagonalizing this
effective dilatation operator will give the spectrum of anomalous dimensions for a class of operators whose
classical dimension is of order $N^2$. Non-planar diagrams have to be included to obtain this dilatation operator.
One consequence of this is that a spin chain description is no longer useful and we have instead passed to a
Cuntz lattice description. In this section we want to answer two questions: 

\begin{itemize}

\item{} The Cuntz lattice description can be employed even before a background is introduced. What is the translation of the
           conserved quantities of the original spin chain into the Cuntz lattice language? 

\item{} Is there any evidence that the effective dilatation operator obtained in the presence of the LLM annulus background is integrable?

\end{itemize}

\subsection{$U_2$ rewritten in the Cuntz Oscillator Language}

Before we obtain the conserved charge $U_2$ in the Cuntz oscillator language, its useful to first write it as a differential operator acting
on gauge invariant operators. We will use this expression when we search for a corresponding conserved charge in the nontrivial background.
It is a straight forward exercise to find that the planar action of 
$$
U_2=\Tr \left((YZZ- ZYZ){d\over dY}{d\over dZ}{d\over d Z}\right)
+\Tr\left((ZZY-YZZ){d\over dZ}{d\over dY}{d\over dZ}\right)
$$
$$
+\Tr\left((ZYZ-ZZY){d\over dZ}{d\over dZ}{d\over dY}\right)
+\Tr\left((ZYY-YZY){d\over dZ}{d\over dY}{d\over dY}\right)
$$
\begin{equation}
+\Tr\left((YYZ-ZYY){d\over dY}{d\over dZ}{d\over dY}\right)
+\Tr\left((YZY-YYZ){d\over dY}{d\over dY}{d\over dZ}\right)\, ,
\end{equation}
on single trace gauge invariant operators matches the action of $U_2$ on the spin chain. To illustrate how to obtain a Cuntz oscillator representation,
consider the first term above: it acts as
\begin{equation} 
ZZY\to YZZ - ZYZ\, .
\end{equation}
This term can be represented in terms of Cuntz oscillators as
\begin{equation} 
\sum_{l=1}^L (a_{l+1}^\dagger a_{l+1}^\dagger -a_l^\dagger a_{l+1}^\dagger)a_la_l\, .
\end{equation}
The second term in the sum should not be truncated to $a_{l+1}^\dagger a_l$ since we have to make sure that there are at least
2 $Z$s in lattice site $l$. The result for $U_2$ is
$$
U_2=\sum_{l=1}^L \Big( (a_{l+1}^\dagger a_{l+1}^\dagger -a_l^\dagger a_{l+1}^\dagger)a_la_l
+(a_{l}^\dagger a_{l}^\dagger -a_{l+1}^\dagger a_{l+1}^\dagger)a_{l+1}a_{l}
$$
$$
+(a_{l}^\dagger a_{l-1}^\dagger -a_{l-1}^\dagger a_{l-1}^\dagger)a_{l}a_{l}
+(a_{l-1}^\dagger -a_{l}^\dagger)a_{l+1}\big[a_l,a_l^\dagger \big]
$$
\begin{equation}
+(a_{l+1}^\dagger -a_{l-1}^\dagger)\big[a_l,a_l^\dagger \big]a_l
+(a_{l}^\dagger -a_{l+1}^\dagger)a_{l-1}\big[a_l,a_l^\dagger \big]\Big)\, .
\end{equation}

Using this procedure it is straight forward to write down the Cuntz oscillator representation for any of the conserved charges of the spin chain.

It is tedious but straight forward to compute the commutator of $U_2$ given above and $D_{\rm 2\, eff}$ - they do not commute. 
It seems natural to consider the operator
\begin{equation} 
U_{\rm 2\, eff}={1\over \chi_B(Z)}U_2 \chi_B(Z)\, .
\end{equation}
It is a simple matter to compute $U_{\rm 2\, eff}$ using the above expression for $U_2$ as a differential operator acting on gauge invariant operators.
Again, $D_{\rm 2\, eff}$ and $U_{\rm 2\, eff}$ do not commute. However, in the leading large $M$ limit the two do commute suggesting 
that this may be an interesting limit of $D_{\rm 2\, eff}$. This is explored in detail in section 3.3 below.

\subsection{Classical Limit}

The original spin chain description of the dilatation operator can be replaced by a sigma model in the limit of a large number of sites.
This sigma model precisely matches the Polyakov action describing the propagation of closed strings in AdS$_5\times$S$^5$, in a particular
limit\cite{Kruczenski:2003gt,Hernandez:2004uw,Bellucci:2004qr}. This has also been extended to other
examples of the AdS/CFT correspondence\cite{Berenstein:2005fa,Frolov:2005ty,Benvenuti:2005cz,deMelloKoch:2005jg,Berenstein:2006qk}.
The Cuntz oscillator description of the dilatation operator is simply an alternative language in the undeformed background; when we consider
the deformed background the spin chain description is not convenient, so that in this case it is best to use the Cuntz oscillator description.
We can obtain a semiclassical description of the Cuntz chain by again considering a large number of sites\cite{Berenstein:2006qk}.
In this subsection we will provide further insight into the relation between the spin chain and Cuntz oscillator descriptions by showing
that in the dual string theory the two descriptions are simply related by a change of worldsheet gauge choice.

The semi-classical limit of the Cuntz chain is obtained by taking $L\sim\sqrt{N}\to\infty$, $\lambda\to\infty$ holding ${\lambda\over L^2}$
fixed and by putting each lattice site into a coherent state (we are discussing the undeformed theory so there are
no negatively occupied sites)
\begin{equation}
|z\rangle = \sqrt{1-|z|^2}\sum_{n=0}^\infty z^n |n\rangle\, ,\qquad |n\rangle = (a^\dagger)^n|0\rangle\, .
\end{equation}
The coherent state parameter of the $l^{th}$ site is traded for a radius and an angle $z_l=r_le^{i\phi_l}$. The action is given, as usual, by
\begin{equation}
S=\int dt\left( i\langle Z|{d\over dt}|Z\rangle - \langle Z|D|Z\rangle\right)\qquad |Z\rangle =\prod_l |z_l\rangle\, .
\end{equation}
After trading the sum over $l$ for an integral over $\sigma$, the action is
\begin{equation}
S=-L\int dt\int_0^1 d\sigma \left( {r^2\dot{\phi}\over 1-r^2}+{\lambda\over L^2}(r'^{2}+r^2\phi'^2)\right)\, . 
\label{CuntzSigma}
\end{equation}
For a detailed derivation the reader could consult \cite{Berenstein:2006qk}. We would like to show how this Cuntz chain sigma model
can be recovered from the standard string sigma model action on $R\times S^3$.

A string moving on $R\times S^3$ can be described by the principal chiral model with su(2) valued currents
\begin{equation}
j_\tau = g^{-1}{\partial g\over\partial \tau}\, ,\qquad j_\sigma = g^{-1}{\partial g\over\partial \sigma}\, ,
\end{equation}
where
\begin{equation}
g=\left [\begin {array}{cc} 
Z &iY\\
\noalign{\medskip}i\bar{Y} &\bar{Z}
\end {array}
\right ]\in {\rm SU(2)}\, ,
\end{equation}
and $Z$ and $Y$ are the coordinates of a sphere
\begin{equation} 
|Z|^2+|Y|^2 = 1\, .
\end{equation}
We are choosing to employ a principal chiral model description since this description manifests the integrability of the model.
Parametrize the sphere coordinates as follows
\begin{equation}
Z = r{e^{i\left (\kappa\tau-\phi\right )}}\, ,\qquad Y = \sqrt {1-{r}^{2}}{e^{i(\varphi+\kappa\tau)}}\, .
\end{equation}
The equations of motion
\begin{equation} 
\partial_\tau j_\tau -\partial_\sigma j_\sigma =0 
\end{equation}
can be obtained from the Polyakov action in conformal gauge and after fixing the residual conformal diffeomorphism freedom by choosing $t=\kappa\tau$.
In this gauge, energy is homogeneously distributed along $\sigma$. To obtain the low energy limit, we take $\kappa\to\infty$ holding $\kappa\dot{r}$,
$\kappa\dot{\phi}$ and $\kappa\dot{\varphi}$ fixed. It is precisely in this gauge and in this limit that\cite{Kruczenski:2003gt} matched the semiclassical
limit of the one loop spin chain to the string sigma model. The Lagrangian in this limit becomes
$$ 
{\cal L} = -{1\over 4}(j_\tau^2-j_\sigma^2)
=r^2\kappa \dot{\phi}-(1-r^2)\kappa \dot{\varphi} + {1\over 2}r^2 \phi'^2+{1\over 2}(1-r^2)\varphi'^2+{1\over 2}{r'^2\over 1-r^2}\, .
$$
The equations of motion following from this action needs to be supplemented by the usual Virasoro constraints, which in this limit are
$ j_\tau^2 + j_\sigma^2=0$ and
\begin{equation}
\kappa \varphi' (1-r^2)^2 +\kappa\phi'r^2 (r^2-1) + O(1) =0\, . 
\label{VC}
\end{equation}
The Cuntz sigma model (\ref{CuntzSigma}) does not contain the field $\varphi$. Thus, it should be eliminated before we can expect to obtain
agreement. Integrate by parts to obtain
\begin{equation}
{\cal L} =r^2\kappa \dot{\phi}- \varphi{\partial\over\partial\tau}(1-r^2)\kappa + {1\over 2}r^2 \phi'^2+{1\over 2}(1-r^2)\varphi'^2+{1\over 2}{r'^2\over 1-r^2}\, .
\end{equation}
Now, using the $\varphi$ equation of motion (to rewrite the coefficients of $\varphi$ in the action) and the (square of the) Virasoro constraint
(\ref{VC}) we find
\begin{equation}
{\cal L}= r^2 \kappa \dot{\phi} + {1\over 2} r^2 \phi'^2+ {1\over 2} {r'^2\over 1-r^2}- {1\over 2} {r^4\over 1-r^2}\phi'^2\, .
\label{PC}
\end{equation}
This does not agree with the result (\ref{CuntzSigma}). 

The disagreement is not surprising: the sigma model (\ref{PC}) is written in a gauge in which energy is distributed homogeneously along the string; the
sigma model (\ref{CuntzSigma}) corresponds to a gauge in which $p_{\varphi}$ (= angular momentum conjugate to $\varphi$) is distributed homogeneously 
along the string. In the Cuntz chain, only $Y$s mark lattice sites; in the usual spin chain both $Y$s and $Z$s mark lattice sites. Consequently to
go from the $\sigma_{cc}$ coordinate of the Cuntz chain to the $\sigma_{sc}$ coordinate of the spin chain, we need to ``add the $Z$s back in''
\begin{equation}
\sigma_{sc}=\sigma_{cc}+\int_0^{\sigma_{cc}} n_z(\sigma') d\sigma'=\sigma_{cc}+\int_0^{\sigma_{cc}} {r^2\over 1-r^2}d\sigma' \, ,\qquad \tau_{sc}=\tau_{cc}\, .
\end{equation}
In this last equation, $n_z(\sigma')=r^2/(1-r^2)$ is the expected number of Cuntz particles (= number of $Z$s) at $\sigma'$. 
It is now straight forward to compute
\begin{equation}
{\partial\sigma_{sc}\over\partial\sigma_{cc}}={1\over 1-r^2},\qquad {\partial\tau_{sc}\over\partial\tau_{cc}}=1,\qquad
{\partial\tau_{sc}\over\partial\sigma_{cc}}=0\, ,
\end{equation}
\begin{equation} 
{\partial\sigma_{sc}\over\partial\tau_{cc}}=-2r^2{\partial\phi\over\partial\sigma_{cc}}=
-2{r^2\over 1-r^2}{\partial\phi\over\partial\sigma_{sc}}\, .
\end{equation}
The integrability condition
\begin{equation}
{\partial\over\partial\tau_{cc}}{\partial\over\partial\sigma_{cc}}\sigma_{sc}
={\partial\over\partial\tau_{cc}}{1\over 1-r^2}
={\partial\over\partial\sigma_{cc}}{\partial\over\partial\tau_{cc}}\sigma_{sc}
={\partial\over\partial\sigma_{cc}}\left( -2r^2{\partial\phi\over\partial\sigma_{cc}}\right)
\end{equation}
is nothing but the $\phi$ equation of motion derived from (\ref{CuntzSigma}). It is now a straight forward exercise 
to verify that after this change of coordinates (\ref{CuntzSigma}) and (\ref{PC}) match perfectly.

The Cuntz oscillator Hamiltonian (\ref{CuntzSigma}) can be written as 
\begin{equation}
S=-L\int dt\int_0^1 d\sigma \left( n_z(\sigma )\dot{\phi}+{\lambda\over L^2}(r'^{2}+r^2\phi'^2)\right)\, . 
\label{gnrslt}
\end{equation}
The advantage of this rewriting is that it holds in general - that is, both for the undeformed and deformed backgrounds.
Inserting the explicit expression for the expected number of Cuntz particles in the deformed background we find
\begin{equation}
S=-L\int dt\int_0^1 d\sigma \left( \left[
{r^2\over 1+{M\over N}-r^2}-{{M\over N}\over {M\over N}-r^2}
\right]\dot{\phi}+{\lambda\over L^2}(r'^{2}+r^2\phi'^2)\right)\, . 
\label{AnnCC}
\end{equation}
{\sl If one drops either of the two terms in square braces one obtains a model which can be related to the low energy limit of the 
principal chiral model and hence is the low energy limit of an integrable model.} The physical interpretation of such a truncation
is clear: keeping only the first term corresponds to focusing on fluctuations localized at the outer edge of the annulus;
keeping only the second term corresponds to focusing on fluctuations localized at the inner edge of the annulus. For these
classes of fluctuations, it seems that the dynamics is integrable. It would be interesting to establish if the full model is
integrable or not.

There is a very natural generalization to multi ring LLM geometries, corresponding to backgrounds created by Schur polynomials
labeled by a Young diagram with more than 4 edges and each edge with a length of $O(N)$. In this case $n_z(\sigma )$ is a sum
of terms, one for each edge. Restricting to fluctuations localized about a particular edge again gives a model which can be 
related to the low energy limit of the principal chiral model and hence is the low energy limit of an integrable model. These
localized excitations have been constructed in \cite{Koch:2008ah}.

The superstar geometry\cite{Myers:2001aq} has been related to an LLM geometry with boundary condition given by a
sequence of concentric alternating black and white rings\cite{Balasubramanian:2005mg}. Rings of the same color have
the same area and the total area of the black rings is $\pi$. As mentioned above, to construct $n_z(\sigma)$ we need to
sum a term for each edge of the multi-ring geometry. We will consider a geometry which corresponds to the Young diagram
shown in figure 1 with $n_1,n_2<<N$. In this case, we sum a very large number of terms and hence may use the Euler-Maclaurin
formula to rewrite the sum as an integral. Carrying this integral out we find (we dropped an additive constant that will
not contribute to the equations of motion)
\begin{equation}
n_z(\sigma) = {\alpha\over\alpha +\beta}{r^2\over 1+{M\over N}-r^2}
\end{equation}
where
\begin{equation}
\alpha ={n_1\over N},\qquad \beta = {n_2\over N}\, . 
\end{equation}

\myfig{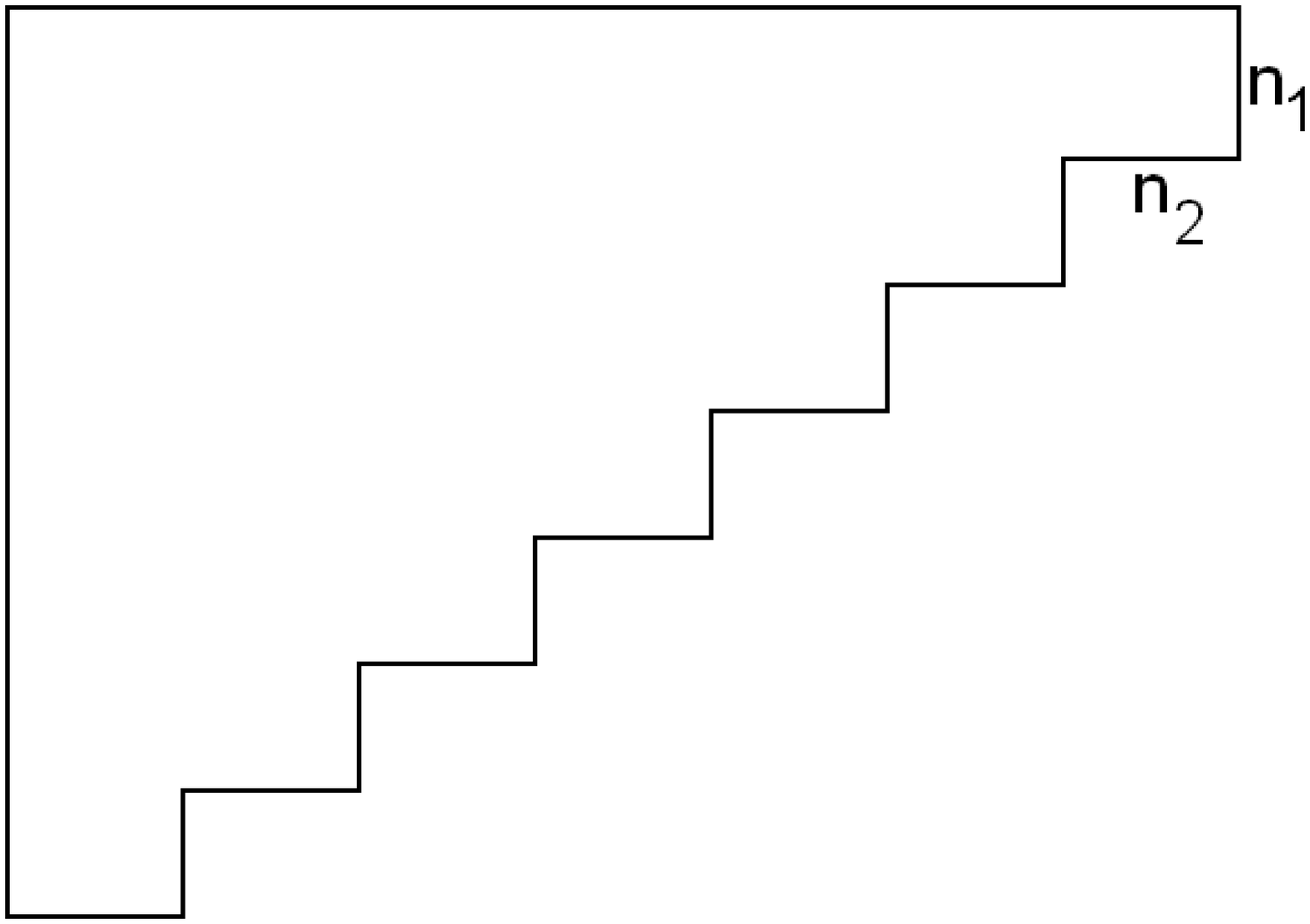}{5.6}{The Young diagram corresponding to the superstar geometry.}

Thus, the semiclassical limit of the Cuntz chain model is
\begin{equation}
S=-L\int dt\int_0^1 d\sigma \left( 
{\alpha\over\alpha +\beta}{r^2\over 1+{M\over N}-r^2}\dot{\phi}+{\lambda\over L^2}(r'^{2}+r^2\phi'^2)\right)\, . 
\end{equation}
After rescaling $t$ we can recover the action (\ref{CuntzSigma}) up to an overall 
multiplicative constant. Thus, the model can again be
related to the low energy limit of an integrable model.

\subsection{Integrability in the Large $M$ Limit}

In this subsection we will consider the large $M$ limit, that is, we take $M,N\to \infty$ and in addition, the ratio ${N\over M}\to 0$.
In this limit we suppress all ${N\over M}$ dependence. The dilatation operator 
for the sector we consider, after subtracting the classical dimension out, can be written as
\begin{equation} 
D= D\left( Z,Y, {d\over dZ}, {d\over dY}\right)\, .
\end{equation}
To get the large $M$ limit of $D_{\rm eff}$ (we denote this operator by $\tilde{D}_{\rm eff}$) we should simply replace ${d\over dZ}$ 
by $MZ^{-1}$ in the above expression to obtain
\begin{equation}
\tilde{D}_{\rm eff}= \tilde{D}_{\rm eff}\left( Z,Y, MZ^{-1}, {d\over dY}\right)\, .
\end{equation}
Expand this operator as
\begin{equation} 
\tilde{D}_{\rm eff}=\sum_{n}\tilde{D}_{\rm eff\, n}
\end{equation}
where $\tilde{D}_{\rm eff\, n}$ has a total of $n$ derivatives with respect to $Y$. From the general structure of a connected planar 
$l$-loop vertex we know that $D$ will act on $l+1$ adjacent sites; thus, it will contain $l+1$ derivatives. The leading contribution 
in the large $M$ limit would come from terms which have all $l+1$ derivatives acting on $Z$ and these are replaced to give an $M^{l+1}\Tr (Z^{-l-1})$.
This leading term is captured in $\tilde{D}_{\rm eff\, 0}$, which up to an arbitrary coefficient, is now determined. 
Since the dimension of $\Tr (Z^J)$ is not corrected, it must be annihilated
by $D$ and hence, in the large $M$ limit it must be annihilated by $\tilde{D}_{\rm eff\, 0}$. This implies that $\tilde{D}_{\rm eff\, 0}$
vanishes. Thus, the leading contribution will in fact come from the $\tilde{D}_{\rm eff\, 1}$. The $l$-loop term will thus have a 
$(g^2 M)^l$ dependence - this is the dependence that the present argument captures\footnote{Of course, our large $M$ limit is a double scaling limit
in which we take $M\to\infty$, $g^2\to 0$ holding $g^2M$ fixed. This limit is the natural one: our effective genus counting parameter is
${1\over M}$ so that $\lambda=g^2M$ is the obvious definition of the 't Hooft coupling. See \cite{Koch:2009jc} for further details.}. 
Since $D$ is dimensionless, and preserves the total number of $Z$s and the total number of $Y$s, it is clear that, to leading order at large $M$
\begin{equation}
\tilde{D}_{\rm eff}=\tilde{D}_{\rm eff\, 1}=\sum_{n}c_n \Tr\left( Z^nYZ^{-n}{d\over dY}\right)
\end{equation}
where the $c_{n}$ depend on $g^2 M$. It is trivial to see that
\begin{equation} 
\left[ \Tr\left( Z^{-n}YZ^n {d\over dY}\right),\Tr\left( Z^{-m}YZ^m {d\over dY}\right)\right]=0\, ,
\end{equation}
so that
\begin{equation}
\left[ \tilde{D}_{\rm eff},\Tr\left( Z^{-m}YZ^m {d\over dY}\right)\right]=0\, ,
\end{equation}
which clearly demonstrates an infinite number of conserved quantities to all loops.

What is the physical meaning of this limit? Recall that the dilatation operator can be read from the two point functions of the theory.
Restricting to operators constructed from scalar fields only is, in general, not possible due to operator mixing. However, it is possible to
show\cite{Beisert:2003tq} that it is consistent to restrict to operators built using only the two complex fields $Y$ and $Z$.
In this case, we can compute the two point functions in a reduced model comprising of only the $Y$ and $Z$ matrices. Interpreted in this way,
the dilatation operator can be understood as implementing the Wick contractions associated with the F-term vertex. For example, consider the
combination
\begin{equation} 
{d\over dZ}+MZ^{-1}
\end{equation}
which replaces ${d\over dZ}$ in transforming the undeformed into the deformed Cuntz chain. The ${d\over dZ}$ term represents a contraction between
the vertex and a $Z^\dagger$ in one of the fields whose two point function we are computing; to see this connection it is useful to remember that
\begin{equation} 
\langle Z_{ij}^\dagger Z_{kl}\rangle = \delta_{jk}\delta_{il}={d\over dZ_{ji}}Z_{kl}\, .
\end{equation}
In contrast to this, the $MZ^{-1}$ term represents a contraction between the vertex and a $Z^\dagger$ in the operator representing the background.
In the large $M$ limit, the contractions with the background completely dominate as compared to contractions with fields belonging
to the operators of the two point functions we are computing. One can think that the matrices entering into the
operators  are ``bits of a string''. In the limit that we consider, the different bits in the string do not
interact with each other - they interact only with the background. We would indeed expect the dynamics to simplify in this limit.

%
%
%
In the large $M$ limit, the action of the Cuntz chain (\ref{AnnCC}) becomes
\begin{equation}
S=-L\int dt\int_0^1 d\sigma \left(
-\dot{\phi}+{\lambda\over L^2}(r'^{2}+r^2\phi'^2)\right)\, .
\end{equation}
Since $\dot{\phi}$ is a total derivative, all time derivatives drop out of the equations of motion. This implies that the dynamics becomes trivial
which is indeed consistent with integrability. It is rather interesting that there is a class of operators in ${\cal N}=4$ super Yang-Mills theory
that have such a simple description.

In terms of the dual LLM boundary condition, the large $M$ limit corresponds to an annulus with a large radius and fixed area, so that the annulus is becoming
very thin.

\section{Discussion}

The problem of computing the anomalous dimensions of operators with a large ($O(N^2)$) {\cal R}-charge corresponds to a generalization, in the dual
gravitational description, to string dynamics in spacetimes that are only asymptotically AdS$_5\times$S$^5$. The problem can again be reduced to
diagonalizing a Hamiltonian. In the AdS$_5\times$S$^5$ spacetime, this Hamiltonian was an integrable spin chain. As a consequence of the fact that
our strings can exchange angular momentum with the background, our Hamiltonian describes Cuntz particles hopping on a lattice. In the gauge theory
description, the terms in the dilatation operator that allow the strings to exchange angular momentum with the background arise from summing (an
infinite number of) non-planar corrections. It is surprisingly straight forward to write down very explicit expressions for the relevant Cuntz chain 
Hamiltonians.

A natural question to ask is if our Cuntz chain Hamiltonians correspond to integrable systems. We don't know. However, we have given some evidence
that the large $M$ limit of our Hamiltonian does admit higher conserved charges and that certain localized semiclassical excitations are described
by the low energy limit of a principal chiral model, so an optimist would indeed conjecture that our Cuntz chain Hamiltonian
is integrable. We hope that we have managed to convince the reader (even if she is pessimistic) that these are interesting limits of the original
${\cal N}=4$ super Yang-Mills theory that warrant further study.

{\vskip 0.5cm}

\noindent
{\it Acknowledgements:} We would like to thank Tom Brown, Kevin Goldstein,
Vishnu Jejjala, Yusuke Kimura, Jeff Murugan and Sanjaye Ramgoolam for enjoyable, 
helpful discussions. 
This work is based upon research supported by the South African Research Chairs Initiative 
of the Department of Science and Technology and National Research Foundation. Any opinion, findings and conclusions 
or recommendations expressed in this material are those of the authors and therefore the NRF and DST do not accept 
any liability with regard thereto. This work is also supported by NRF grant number Gun 2047219.

\appendix

\section{Two Point Functions}

In this Appendix we will compute the two point functions used in section 2. We only want these two point functions to the leading
order in an $M+N$ expansion. We will use $f_B$ to denote the product of the weights of Young diagram $B$. It is straight forward
to obtain
\begin{equation} 
f_B ={G_2 (N+M+1)\over G_2 (N+1)G_2 (M+1)} \, ,
\end{equation}
where $G_2 (n+1)$ is the Barnes function defined by ($\Gamma (z)$ is the Gamma function)
\begin{equation} 
G_2 (z+1)=\Gamma (z)G_2 (z)\, .
\end{equation}
In particular, for an integer $z=n$ we have
\begin{equation} 
G_2 (n+1)=\prod_{k=1}^{n-1} k!\, .
\end{equation}

First consider the free field theory two point function
\begin{equation} 
\left\langle \chi_B(Z)\chi_B(Z^\dagger) \Tr (Y^2 Z^{J_0})\Tr (Y^2 Z^{J_0})^\dagger\right\rangle .
\end{equation}
According to \cite{Koch:2008cm}, this two point function is given, at the leading order in a large $M+N$ expansion, by $f_B$ times 
$\left({M+N\over N}\right)^{J_0}$ times the free field theory two point function in the trivial background
\begin{equation}
\left\langle\Tr (Y^2 Z^{J_0})\Tr (Y^2 Z^{J_0})^\dagger\right\rangle =N^{J_0+2}+O(N^{J_0})\, .
\end{equation}
Thus,
\begin{equation} 
\left\langle \chi_B(Z)\chi_B(Z^\dagger) \Tr (Y^2 Z^{J_0})\Tr (Y^2 Z^{J_0})^\dagger\right\rangle = f_B (N+M)^{J_0}N^2\, .
\end{equation}
Arguing in exactly the same way, we find
\begin{equation} 
\left\langle \chi_B(Z)\chi_B(Z^\dagger) \Tr (YZ^n Y Z^{J_0-n})\Tr (YZ^m Y Z^{J_0-m})^\dagger\right\rangle = f_B (N+M)^{J_0}N^2\delta_{mn}\, .
\end{equation}

Next, consider
\begin{equation} 
\left\langle \chi^{(1)}_{B,B'}(Z,YZ^{J_0+1}Y)\chi^{(1)}_{B,B'}(Z,YZ^{J_0+1}Y)^\dagger\right\rangle 
\end{equation}
where
\begin{equation}
\chi^{(1)}_{B,B'}(Z,YZ^{J_0+1}Y)=\left( YZ^{J_0+1}Y\right)_{ij}{\partial\over\partial Z_{ij}}\chi_B(Z)\, .
\end{equation}
These correlators have been computed in \cite{deMelloKoch:2007uu}. The result is
\begin{equation} 
\left\langle \chi_{B,B'}^{(1)}(Z,W)\chi_{B,B'}^{(1)}(Z^\dagger ,W^\dagger)\right\rangle = 
      {{\rm hooks}_B\over {\rm hooks}_{B'}} f_B F_0 + c_{BB'} f_B F_1 \, . 
\end{equation}
We find $c_{BB'}=M$ and ${{\rm hooks}_B\over {\rm hooks}_{B'}}=MN$. Further, for the open string word $W=YZ^{J_0+1}Y$ we find that
at the leading order
\begin{equation} 
F_0=N^{J_0+2}\left({M+N\over N}\right)^{J_0+1}\, ,\qquad F_1\sim N^{J_0-1}\left({M+N\over N}\right)^{J_0+1}\, ,
\end{equation}
so that, to the leading order we have
\begin{equation} 
\left\langle \chi^{(1)}_{B,B'}(Z,YZ^{J_0+1}Y)\chi^{(1)}_{B,B'}(Z,YZ^{J_0+1}Y)^\dagger\right\rangle =f_BMN^2 (M+N)^{J_0+1}\, .
\end{equation}
This correlator result can also be written as
\begin{equation} 
\left\langle \chi_B(Z)\chi_B(Z^\dagger) \Tr (YZ^{-1} Y Z^{J_0+1})\Tr (YZ^{-1} Y Z^{J_0+1})^\dagger\right\rangle = f_B {1\over M}(N+M)^{J_0+1}N^2 \, .
\end{equation}
Again arguing in exactly the same way, we find
\begin{equation} 
\left\langle \chi_B(Z)\chi_B(Z^\dagger) \Tr (YZ^{-n} Y Z^{J_0+n})\Tr (YZ^{-m} Y Z^{J_0+m})^\dagger\right\rangle = f_B {1\over M^n}(N+M)^{J_0+n}N^2\delta_{mn}\, .
\end{equation}
Notice that at large enough $M$ that ${N\over M}$ can be neglected, all the gauge invariant operators considered have exactly the same two point function.

Finally, by using the methods of this Appendix, we can obtain the general result
\begin{equation} 
\langle {\cal O}_B(\{ p\}, J_0){\cal O}_B(\{ p\}, J_0)^\dagger\rangle = N^2 f_B {(M+N)^{J_0+p_-}\over M^{p_-}}\, ,
\end{equation}
where $\{p\}$ denotes the occupation numbers of the Cuntz chain and $p_-$ is negative the sum of all the negative occupation numbers.
Thus, we have the correspondence
\begin{equation} 
{\cal O}_B(\{ p\}, J_0)\leftrightarrow\sqrt{N^2 f_B {(M+N)^{J_0+p_-}\over M^{p_-}}}|\{ p\}\rangle \, ,
\end{equation}
between operators and normalized Cuntz lattice states.

\section{More on the Cuntz Chain}

To specify the general Cuntz chain model (\ref{gnrslt}) one needs to specify the expected number of Cuntz particles $n_z(\sigma)$. Given $n_z(\sigma)$,
what is the corresponding supergravity background? Using the results of the first of \cite{Chen:2007gh} as well as (\ref{gnrslt}), the metric on the $y=0$
plane and the circle along which the string moves (parametrized by\footnote{The angular momentum along this circle is due to the $Y$ fields appearing
in the gauge invariant operator dual to the string.} $\varphi$) can be written as
\begin{equation} 
ds^2 =-h^{-2}(Dt)^2 + h^2 dzd\bar{z}+h^{-2}d\varphi^2\, , \qquad Dt = dt -{1\over 2}i\bar{V}dz +{1\over 2}iVd\bar{z}\, , 
\end{equation}
\begin{equation} 
V={n_z\over\bar{z}}, \qquad h^4={\partial V\over\partial z},\qquad z=re^{i\phi}\, .
\end{equation}

\section{Explicit Expressions for the Two Loop Dilatation Operator}

In these expressions hatted indices are again to be dropped,
$\theta(x)=1$ if $x>0$ and vanishes otherwise. To obtain this result we have assumed $J_0>0$, $J_0-p\ge 0$ and 
$p\ge 0$ - assumptions which can easily be relaxed if need be.
$$
\delta D_{\rm 4,0\,\, eff}{\cal O}_p^{J_0;J_1,...,J_k}=
4({\cal O}_{1}^{J_0;J_1,...,J_k}-{\cal O}_0^{J_0;J_1,...,J_k})\times
$$
\begin{equation}
\times N(N+M)(\delta_{p=0}+\delta_{p=J_0}-\delta_{p=1}-\delta_{p=J_0-1})\, ,
\end{equation}
{\vskip 0.1cm}
$$\delta D_{\rm 4,+\,\, eff}{\cal O}_p^{J_0;J_1,...,J_k}=4(M+2N)\theta(p-1)\times$$
$$\times({\cal O}_0^{J_0-p+1;J_1,...,J_k,p-1}-{\cal O}_1^{J_0-p+1;J_1,...,J_k,p-1})
$$
$$
+4(M+2N)\theta (J_0-p-1)({\cal O}_0^{p+1;J_1,...,J_k,J_0-p-1}-{\cal O}_1^{p+1;J_1,...,J_k,J_0-p-1})
$$
$$
+4(N+M)\theta (p)({\cal O}_1^{J_0-p;J_1,...,J_k,p}
-{\cal O}_0^{J_0-p;J_1,...,J_k,p})
$$
$$
+4(N+M)\theta (J_0-p)({\cal O}_1^{p;J_1,...,J_k,J_0-p}
-{\cal O}_0^{p;J_1,...,J_k,J_0-p})
$$
\begin{equation}
+\sum_{s=1}^{J_0-1}4N(\delta_{p=0}+\delta_{p=J_0})({\cal O}_1^{J_0-s;J_1,...,J_k,s}
-{\cal O}_0^{J_0-s;J_1,...,J_k,s})\, ,
\end{equation}
{\vskip 0.1cm}
$$\delta D_{\rm 4,-\,\, eff}{\cal O}_p^{J_0;J_1,...,J_k}=
4N\sum_{j=1}^k \, J_j\, (\delta_{p=0}+\delta_{p=J_0})\times$$
$$\times ({\cal O}_1^{J_0+J_j;J_1,...,\hat{J}_j,...,J_k}
-{\cal O}_0^{J_0+J_j;J_1,...,\hat{J}_j,...,J_k})$$
\begin{equation}
-4N(\delta_{p=0}+\delta_{p=J_0})\sum_{i=1}^k \delta_{J_i=1}
({\cal O}_1^{J_0+1;J_1,...,\hat{J}_i,...,J_k}
-
{\cal O}_0^{J_0+1;J_1,...,\hat{J}_i,...,J_k})\, ,
\end{equation}
{\vskip 0.1cm}
$$\delta D_{\rm 4,+-\,\, eff}{\cal O}_p^{J_0;J_1,...,J_k}=4\theta(p)\sum_{i=1}^k
J_i({\cal O}_1^{J_0+J_i-p;J_1,...,\hat{J}_i,...,J_k,p}
$$
$$
-{\cal O}_0^{J_0+J_i-p;J_1,...,\hat{J}_i,...,J_k,p})+4\theta (J_0-p)
\sum_{i=1}^k
J_i({\cal O}_1^{J_i+p;J_1,...,\hat{J}_i,...,J_k,J_0-p}
$$
$$
-{\cal O}_0^{J_i+p;J_1,...,\hat{J}_i,...,J_k,J_0-p})-4\theta (J_i-1)\sum_{i=1}^k J_i
({\cal O}_1^{p+1;J_1,...,\hat{J}_i,...,J_k,J_0+J_i-p-1}
$$
$$
-{\cal O}_0^{p+1;J_1,...,\hat{J}_i,...,J_k,J_0+J_i-p-1})
-4\theta (J_i-1)\sum_{i=1}^k J_i ({\cal O}_1^{J_0-p+1;J_1,...,\hat{J}_i,...,J_k,J_i+p-1}
$$
\begin{equation}
-{\cal O}_0^{J_0-p+1;J_1,...,\hat{J}_i,...,J_k,J_i+p-1})\, ,
\end{equation}
{\vskip 0.1cm}
$$\delta D_{\rm 4,++\,\, eff}{\cal O}_p^{J_0;J_1,...,J_k}=
4\sum_{r=1}^{p-2}({\cal O}_0^{J_0-p+1;J_1,...,J_k,r,p-r-1}$$
$$-{\cal O}_1^{J_0-p+1;J_1,...,J_k,r,p-r-1})+4\sum_{r=1}^{J_0-p-2}
({\cal O}_0^{p+1;J_1,...,J_k,r,J_0-p-r-1}
$$
$$
-{\cal O}_1^{p+1;J_1,...,J_k,r,J_0-p-r-1})+4\theta(p)\sum_{s=1}^{J_0-p-1}(
{\cal O}_1^{J_0-p-s;J_1,...,J_k,s,p}
$$
$$ -{\cal O}_0^{J_0-p-s;J_1,...,J_k,s,p})+4\theta(J_0-p)\sum_{s=1}^{p-1}
({\cal O}_1^{p-s;J_1,...,J_k,s,J_0-p}
$$
\begin{equation}
-{\cal O}_0^{p-s;J_1,...,J_k,s,J_0-p})\, ,
\end{equation}
{\vskip 0.1cm}
\begin{equation}
\delta D_{\rm 4,--\,\, eff}{\cal O}_p^{J_0;J_1,...,J_k}=0\, .
\end{equation}
Setting $M=0$ in the above expressions gives exact agreement with Appendix E of \cite{Beisert:2003tq} except for the last term in
our expression for $\delta D_{\rm 4,-}$. The extra term that we have ensures that no joinings between the trace with the $Y$s and
a trace without $Y$s and a single $Z$ can occur. In this case, the ${d\over dZ_{ij}}$ in $\delta D_{\rm 4\,\, eff}$ acts on $\Tr (Z)$ to produce 
$\delta_{ij}$. This vanishes because ${d\over dZ_{ij}}$ appears inside a commutator; the extra term is needed.

\end{document}